\documentclass[letterpaper, 10 pt, conference]{ieeeconf}  

\IEEEoverridecommandlockouts
\makeatletter
\let\NAT@parse\undefined
\makeatother

\usepackage{amsmath}
\usepackage{amssymb,bm}
\usepackage{amsfonts}
\usepackage{enumerate}
\usepackage{tabulary}
\usepackage{multirow}
\usepackage{cite}
\usepackage{lipsum} 
\usepackage{verbatim}
\usepackage[hidelinks]{hyperref}
\usepackage{url}
\usepackage{balance}
\usepackage{graphicx}
\usepackage{subcaption}
\usepackage[font=footnotesize]{caption}
\usepackage{epstopdf}
\usepackage[misc]{ifsym}
\usepackage{color}
\usepackage{xcolor}
\usepackage{xxcolor}
\usepackage{tikz}
\usepackage{pgf}
\usepackage{pgfplots}
\usepackage{float}
\usepackage{siunitx}
\usepackage{ifthen}
\usepackage{forloop}
\usepackage{listings}
\usepackage{booktabs}
\usepackage[lined,boxed,commentsnumbered,linesnumbered]{algorithm2e}

\usepackage[utf8]{inputenc}
\usepackage{pgfplots}
\DeclareUnicodeCharacter{2212}{−}
\usepgfplotslibrary{groupplots,dateplot}
\usetikzlibrary{patterns,shapes.arrows}
\pgfplotsset{compat=newest}

\hypersetup{
  colorlinks    = true, 
  urlcolor      = blue, 
  linkcolor     = blue, 
  citecolor     = blue   
}
\definecolor{codegreen}{rgb}{0,0.6,0}
\definecolor{codepurple}{rgb}{0.58,0,0.82}
\definecolor{backcolour}{rgb}{0.95,0.95,0.92}
\lstdefinestyle{buzz}{
    backgroundcolor=\color{black!5},   
    commentstyle=\color{codegreen},
    keywordstyle=\color{blue},
    numberstyle=\tiny\color{black!30},
    stringstyle=\color{codepurple},
    basicstyle=\footnotesize\ttfamily,
    breakatwhitespace=false,         
    breaklines=true,                 
    captionpos=b,                    
    keepspaces=true,                 
    numbers=left,                    
    numbersep=5pt,                  
    showspaces=false,                
    showstringspaces=false,
    showtabs=false,                  
    tabsize=2,
}
\renewcommand{\vec}[1]{#1}

\newcommand{\x}{x}
\renewcommand{\u}{u}

\renewcommand{\u}{\vec{u}}

\newcommand{\set}[1]{\mathbb{#1}}
\newcommand{\R}{\mathbb{R}}

\newcommand{\K}{\mathcal{K}}

\usepackage{stackengine}

\newtheorem{corollary}{Corollary}

\newtheorem{definition}{Definition}

\newtheorem{example}{Example}
\newtheorem{insight}{Insight}

\pgfdeclarelayer{foreground}
\pgfsetlayers{main,foreground}

\title{\LARGE \bf
Addressing Relative Degree Issues in Control Barrier Function Synthesis with Physics-Informed Neural Networks
}

\author{Lukas Brunke$^1$, Siqi Zhou$^1$, Francesco D'Orazio$^2$, and Angela P. Schoellig$^1$
\thanks{%
$^1$Learning Systems and Robotics Lab and the Munich Institute of Robotics and Machine Intelligence, Technical University of Munich, 80333 Munich, Germany. 
{Email: \tt\small firstname.lastname@tum.de}}
\thanks{$^2$Department of Computer, Control and Management Engineering, Sapienza University of Rome, Rome, Italy. {E-mail: \tt\small lastname@diag.uniroma1.it}.}
}

\begin{document}

\maketitle
\thispagestyle{empty}
\pagestyle{empty}

\begin{abstract}
In robotics, control barrier function (CBF)--based safety filters are commonly used to enforce state constraints. A critical challenge arises when the relative degree of the CBF varies across the state space. This variability can create regions within the safe set where the control input becomes unconstrained. When implemented as a safety filter, this may result in chattering near the safety boundary and ultimately compromise system safety. To address this issue, we propose a novel approach for CBF synthesis by formulating it as solving a set of boundary value problems. The solutions to the boundary value problems are determined using physics-informed neural networks~(PINNs). Our approach ensures that the synthesized CBFs maintain a constant relative degree across the set of admissible states, thereby preventing unconstrained control scenarios. We illustrate the approach in simulation and further verify it through real-world quadrotor experiments, demonstrating its effectiveness in preserving desired system safety properties.
\end{abstract}

\section{INTRODUCTION}

\begin{figure}
    \centering
    \begin{subfigure}{\columnwidth}
        \centering
        \includegraphics[width=\textwidth]{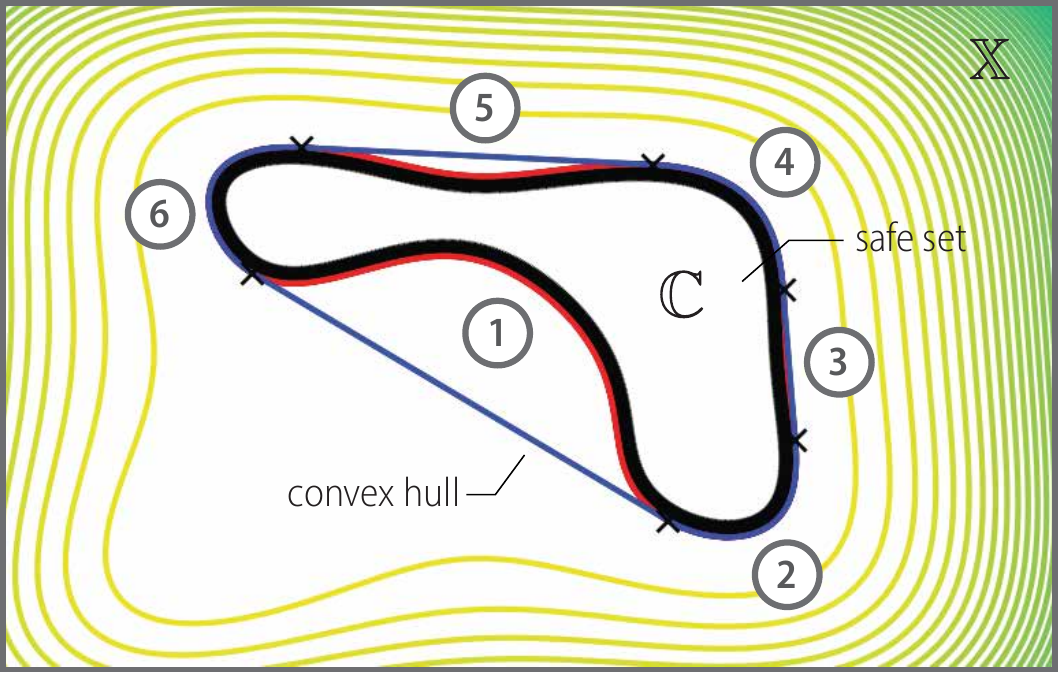}
        \caption{Compact safe set.}
    \end{subfigure}
    \begin{subfigure}{\columnwidth}
        \centering
        \includegraphics[width=\textwidth]{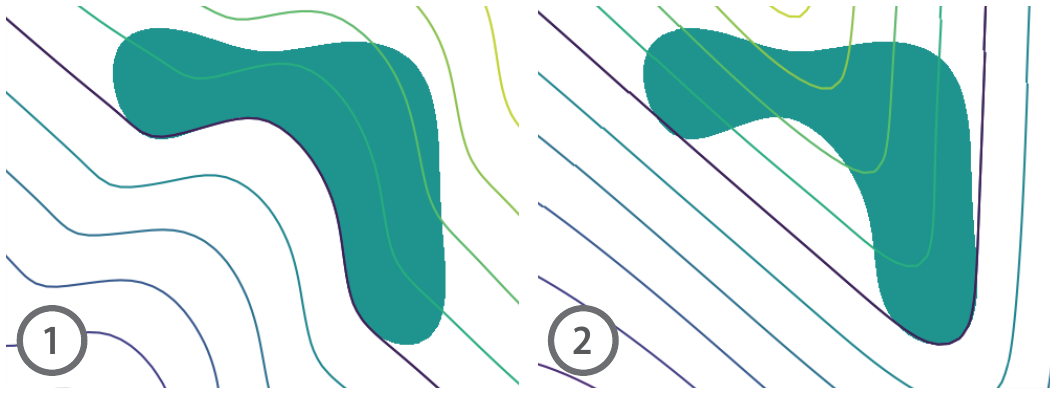}
        \caption{Level sets of two CBFs covering different segments of the safe set boundary.}
    \end{subfigure}
    \caption{An illustration of our proposed approach where we leverage multiple control barrier functions (CBFs) to mitigate the varying relative degree issue in certifying compact safe sets~\cite{brunke2024practical}. In this work, we introduce an alternative perspective by formulating CBF synthesis as boundary value problems, which are solved using physics-informed neural networks (PINNs). This approach allows us to mitigate the relative degree issue without conservative safe set approximations. (a)~As an example, a non-convex, compact safe set is parameterized by multiple CBFs, each covering a segment of the safe set boundary. (b)~The level sets of two representative CBFs are shown. The flexibility of PINNs allows us to closely approximate the original safe set boundary for both convex segments and non-convex segments (highlighted in red between two crosses in the top panel).
    }
    \label{fig:bvp_illustration}
\end{figure}

In robotics, safety filters are gaining increasing attention as a means of providing safety guarantees to learning-based control methods that are not inherently designed to be safe~\cite{DSL2021}. A common approach to safety filter design involves the use of control barrier functions (CBFs)~\cite{hsu2023safety,ames2019a,wieland2007constructive}. When augmenting a system with a CBF-based safety filter, the objective is to certify or minimally adjust the control commands computed by an otherwise unsafe policy. This idea has been applied across a wide range of robotic problems, including, but not limited to, manipulation, locomotion, as well as autonomous driving and flight.

In the CBF safety filter literature, while both continuous-time and discrete-time implementations have been proposed, it is common to use a continuous-time formulation of CBFs~\cite{ames2019a,zeng2021safety}. This is due to the fact that, for control-affine systems, a continuous-time formulation leads to a quadratic program (QP) that can be solved efficiently online~\cite{hsu2023safety,ames2019a}. However, a subtle but important issue often overlooked in this setting is the problem of varying relative degrees. The relative degree determines the number of times the barrier function must be differentiated before the control input appears explicitly. If this property is not properly checked, the resulting safety filter may become inactive in the safe set. When such points lie close to the safe set boundary, this can lead to chattering or even constraint violations~\cite{brunke2024practical}. Notably, this issue can arise even in the simple case of linear systems with quadratic CBFs.

In recent work, several approaches have been proposed to address this issue. Some of the approaches include using multiple CBFs with affine or quadratic forms~\cite{brunke2024practical,brunke2024preventingunconst, black2023adaptation}, reformulating the safety filter QP to relax the assumption on uniform relative degree~\cite{jankovic2018robust}, introducing penalization terms to numerically mitigate chattering~\cite{brunke2024practical}, and reconstructing CBFs such that the inactivity issue does not compromise safety during task execution~\cite{tan2021high,brunke2024practical}. While effective, these approaches often rely on conservative approximations of the safe set or only address the problem in an ad-hoc manner~(i.e., a CBF is first constructed and then verified or modified retrospectively to ensure proper behaviour).

In this work, we propose an alternative perspective on CBF synthesis. In particular, we argue that the synthesis of CBFs should begin with the design of their gradients. By constructing CBFs from their gradients, we can directly mitigate issues such as varying relative degrees. Since CBFs inherently involve boundary conditions that are meant to capture the geometry of the safe set, the problem of CBF synthesis naturally leads to solving boundary value problems. Solving such problems is non-trivial, especially for generic control-affine systems; to address this, we leverage physics-informed neural networks (PINNs)~\cite{raissi2019physics,baty2024hands} as a tool for solving the associated boundary value problems through supervised learning. This overall approach enables us to mitigate the varying relative degree issue without relying on retrospective modifications or conservative safe set approximations (\autoref{fig:bvp_illustration}). As compared to existing CBF synthesis methods such as sum-of-squares (SOS) approaches~\cite{dai2022convex}, learning-based methods~\cite{qin2022sablas,taylor2020a,chen2024learning}, and Hamilton-Jacobi (HJ) reachability-based techniques~\cite{choi2021robust,tonkens2022refining}, our approach specifically focuses on addressing the varying relative degree issue through gradient-based design.

Our contributions are as follows:
\begin{enumerate}
    \item We introduce a new perspective on addressing the varying relative degree issue by treating the problem of CBF synthesis as solving boundary value problems.
    \item We propose a PINN-based method that allows us to synthesize CBFs that are free from inactivity issues without relying on retrospective modifications or conservative approximations.
    \item We demonstrate the effectiveness of our approach in both simulation and real-time quadrotor experiments.
\end{enumerate}
\section{PROBLEM FORMULATION}
\begin{figure}
    \centering
    \includegraphics[width=\columnwidth]{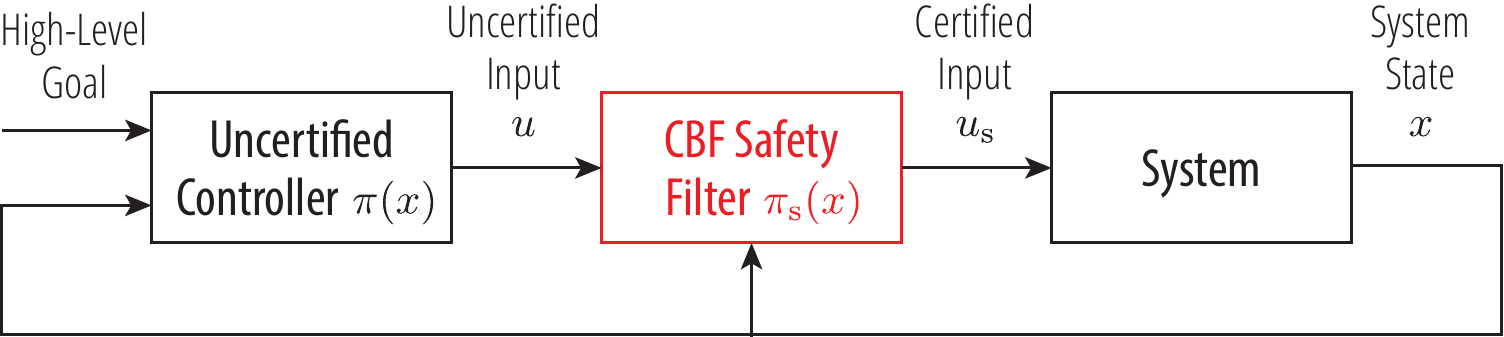}
    \vspace{1em}
    \caption{A block diagram of a typical safety filter control architecture. Given an uncertified policy $\pi(x)$, a safety filter $\pi_\text{s}(x)$ is designed to safeguard the system by making minimal adjustments to the control inputs when they are deemed unsafe.}
    \label{fig:blockdiagram}
\end{figure}

We consider the control architecture shown in \autoref{fig:blockdiagram} and systems with continuous-time control affine dynamics:
\begin{equation}
\label{eqn:nonlinear_affine_control}
    \dot{x} = f(x) + g(x) u,
\end{equation}
where $x\in\set{X}\subset\set{R}^n$ is the state of the system with $\set{X}$ denoting the set of admissible states, $u\in\set{R}^m$ is the input of the system, and $f:\set{R}^n\mapsto \set{R}^n$ and $g:\set{R}^n\mapsto \set{R}^{n\times m}$ are locally Lipschitz continuous functions.

 Given an uncertified policy $\pi(x)$, our goal is to design a safety filter to safeguard the system---ensuring that the state of the system remains within a given safe set~$\set{C}\subseteq \set{X}$. The safe set $\set{C}$ is assumed to be compact and is parameterized as the zero-superlevel set of a continuously differentiable function $h:\set{X}\mapsto\set{R}$:
\begin{equation*}
    \set{C} = \{x\in\set{X}\:\:|\:\: h(x)\ge 0\}.
\end{equation*}
The boundary of the safe set is $\partial \set{C}=\{x\in\set{X}\:\:|\:\: h(x)=0\}$ with $\nabla h(x) \neq 0$ for all $x\in\partial \set{C}$, and the interior of the safe set is $\text{Int}(\set{C}) =\{x\in\set{X}\:\:|\:\: h(x)>0\}$.

\section{BACKGROUND}
In this section, we provide the background on CBF-based safety filters to facilitate our discussion. We begin by introducing our notations and necessary definitions.

We denote the Euclidean norm as $\lVert \cdot \rVert$. A set of consecutive integers is denoted by $\set{Z}_{[a, b]}$, where $a$ and $b$ are positive integers with $a \leq b$. The notations $\text{Int}(\cdot)$ and $\text{Conv}(\cdot)$ correspond to the interior and the convex hull of a set, respectively. The Lie derivative of a function $h$ along a vector field $f$ is written as $L_f h$, and $\nabla$ represents the gradient operator. The symbol $\circ$ denotes function composition.

\begin{definition}[Forward Invariant Set] Consider a system with dynamics $\dot{x}=f(x)$ with $f:\set{X}\mapsto\set{X}$ being a Lipschitz continuous function. A set ${\set{C}\subseteq\set{X}}$ is a forward invariant set for the system if $x(0) \in\set{C}$ implies $\x(t)\in\set{C}$ for all $t\in \set{T}_{\x_0}^+$, where $\set{T}_{\x_0}^+$ is the maximum time interval of existence for the state trajectory initialized at $x(0)=x_0$. 
\end{definition}

\begin{definition}[Extended Class-$\K$ Function]
\label{def:kappa_inf_extended}
A continuous function $\gamma:\set{R}\mapsto\set{R}$ is an extended class-$\K$ function, denoted by $\K_e$, if it is strictly increasing and passes through the origin~(i.e., $\gamma(0)=0$).
\end{definition}

\begin{definition}[Relative Degree~\cite{ames2019a}]
\label{def:rel-degree}
Let $h:\set{X}\mapsto \set{R}$ be a $\rho^{\text{th}}$-order differentiable function. A system with dynamics~\eqref{eqn:nonlinear_affine_control} and output equation~$y=h(x)$ is said to have a relative degree of $\rho$ if $\rho$ is the smallest integer between 1 and $n$ such that $L_g L_f^{\rho - 1} h(x) \neq 0$ for all $x \in \set{X}$.
\end{definition}

In this work, we consider systems with dynamics as specified in~\eqref{eqn:nonlinear_affine_control} and say that a $\rho^{\text{th}}$-order differentiable function, $h(x)$, has a relative degree of $\rho$ if the system with~\eqref{eqn:nonlinear_affine_control} and output equation~$y=h(x)$ has a relative degree of $\rho$.

\begin{definition}[CBF~\cite{ames2019a}]
\label{def:cbf}
Let $\set{C} \subseteq \set{X}$ be the zero-superlevel set of a continuously differentiable function $h: \set{X} \to \R$ with $\partial \set{C} = \{x\in\set{X}\:\:|\:\: h(x) = 0\}$ and $\nabla_x h(x) \neq 0 $ for all $x\in\partial \set{C}$. The function $h$ is a CBF for~\eqref{eqn:nonlinear_affine_control} if there exists a function $\gamma\in\K_e$ such that the following condition is satisfied:
\begin{equation*}
\label{eqn:cbf_lie_derivative}
\max_{\u \in \R^m}\hspace{0.5em}\left[L_\vec{f} h(\x) + L_\vec{g} h(\x) \u \right] \geq - \gamma(h(x)),\quad \forall\x \in \set{X}.
\end{equation*}
\end{definition}

Based on \autoref{def:cbf}, we can define a set of certified inputs for every $x\in\set{X}$:
\begin{equation*}
    \label{eqn:cbf_input_set}
 \set{U}_{\text{cbf}}(x) = \{u\in\R^m \:\vert \:  \dot{h}(x,u) \geq - \gamma(h(x))\} ,
\end{equation*}
where $\dot{h}(x,u) = L_\vec{f} h(\x) + L_\vec{g} h(\x) \u$.

As discussed in~\cite{ames2019a}, CBFs can be used to verify the forward invariance of a safe set under the closed-loop dynamics of a system.
\begin{corollary}[Forward Invariance of Safe Set~\cite{ames2019a}]\label{cor:theorem_2_ames}
    Let $\set{C} \subset \R^n$ be a set defined as the superlevel set of a continuously differentiable function $h: \set{X} \subset \R^n \mapsto \R$. If $h$ is a control barrier function on $\set{X}$ and $\frac{\partial h}{\partial \x} (\x) \neq 0$ for all $\x \in \partial \set{C}$, then any Lipschitz continuous policy $\pi(\x) \in \set{U}_{\text{cbf}}(\x)$ for the system~\eqref{eqn:nonlinear_affine_control} renders the set $\set{C}$ safe. 
\end{corollary}

If the policy $\pi:\set{X}\mapsto\set{R}^m$ is initially not designed to be safe (i.e., $\pi(x)\notin\set{U}_\text{cbf}(x)$ for some $x\in\set{X}$), a QP is often formulated as a safety filter to modify potential unsafe inputs~\cite{ames2019a}:
\begin{subequations}
\label{eqn:cbf_qp}
\begin{align}
\pi_\text{s}(x)=\arg\min_u &\:\:||u - \pi(x)||^2\\
    \text{s.t.}&\:\: L_fh(x) + L_gh(x) u \ge -\gamma(h(x)).
\end{align}
\end{subequations}
Intuitively, the safety filter finds an input that best matches the one computed by $\pi$ while satisfying the CBF condition to render the safe set $\set{C}$ positive invariant.

\section{RELATIVE DEGREE ISSUES IN CBF SYNTHESIS}
\label{sec:relative_degree}
To apply the safety filter formulation in~\eqref{eqn:cbf_qp}, we, in fact, require the CBF to have a relative degree of one for all $x\in\set{X}$. However, for compact safe sets, a single continuously differentiable function would not satisfy the relative degree condition. In particular, there would be points in the safe set where $L_g h(x)=0$. Identifying the relative degree of a CBF is not always straightforward, especially in the case of generic nonlinear control affine dynamics of the form~\eqref{eqn:nonlinear_affine_control}. 

We illustrate the varying relative degree issue using a simple example below and generalize the conclusion to systems with nonlinear control affine dynamics. 

\begin{example}[Standard CBF Parametrization]
\label{ex:standard_cbf}
      Consider a single integrator system $\dot{x}  = u$ with a compact constraint set $\mathbb{C}_\text{1d} = \{x\in\set{X}\:\:|\:\: x_\text{min}\le x \le x_\text{max}\}$ with $x_\text{max}> x_\text{min}$. A standard form of CBF often seen in the literature is a quadratic function. For our example, one possible quadratic CBF is $h(x) = (x-x_\text{min})(x_\text{max}-x)$. While the CBF satisfies the set of conditions in~\autoref{def:cbf}, its gradient vanishes at $x_\text{zero} = (x_\text{min}+x_\text{max})/2$. At this point, we have $L_gh(x_\text{zero})=0$; the safety filter QP in~\eqref{eqn:cbf_qp} is unconstrained. In general, by the mean value theorem, one can show that for any continuously differentiable function $h(x)$, there exists a point $ x_\text{zero}$ in the interior of the safe set such that $\nabla h(x_\text{zero}) = \big(h(x_\text{max})-h(x_\text{min})\big) / (x_\text{max} - x_\text{min}) = 0$. In other words, there exists at least one point in the safe set, where the safety filter in~\eqref{eqn:cbf_qp} will be inactive. 
\end{example}

This example shows that using a single continuously differentiable function to parametrize a compact safe set is not sufficient. For a linear system with quadratic CBFs, when $g(x)$ does not have full row rank, the set of states corresponding to a non-unitary relative degree is a hyperplane cutting through the safe set; enforcing the CBF constraint close to the point where the hyperplane intersects the safe set is problematic. This conclusion applies to general compact safe sets parameterized by a single continuously differentiable CBF.

\begin{insight}[CBF Over Compact Sets]
\label{insight:cbf_over_compact_sets}
Consider dynamics in~\eqref{eqn:nonlinear_affine_control} and a compact set $\set{C}\subseteq \set{X}$ that is parameterized as the zero-superlevel set of a continuously differentiable function $h:\set{X}\mapsto\set{R}$ with $\partial \set{C} = \{x\in\set{X}\:\:|\:\: h(x) = 0\}$ and $\text{Int}(\set{C}) = \{x\in\set{X}\:\:|\:\: h(x) > 0\}$. The gradient satisfies $\nabla h(x) \neq 0 $ for all $x\in\partial \set{C}$. There exists at least one point in $\text{Int}(\set{C})$, where $\nabla h(x)=0$; at such points, the Lie derivative $L_gh(x)$ is zero, and the input $u$ is unconstrained in the safety filter~\eqref{eqn:cbf_qp}. Moreover, for cases where the dimension of the state space is $n\ge 2$, if the $i$-th column of $g(x)$, denoted by $g_i(x)$, is a non-zero constant vector over $\set{X}$, there exists at least one point in $\partial \set{C}$, where $\nabla h(x)$ is orthogonal to $g_i(x)$; at these points, $L_{g_i}h(x)=0$, and the $i$-th element of the input $u_i$ is unconstrained in the safety filter~\eqref{eqn:cbf_qp}.
\end{insight}

\begin{proof}
Since $h$ is continuous over the compact set~$\set{C}$, by the extreme value theorem, there exists a point $x^* \in\set{C}$, where the function $h$ attains its maximum value. Furthermore, since $\set{C}$ is a compact zero-superlevel set of a continuously differentiable function $h$ and $h(x)>0$ for $x\in\text{Int}(\set{C})$, the maxima must lie in the interior of the set $\text{Int}(\set{C})$. By the interior extremum theorem, a maximum of $h$ is a stationary point with zero gradients (i.e., $\nabla h(x)=0$)~\cite{bertsekas1997nonlinear}. Thus, there exists at least one point in $\text{Int}(\set{C})$ where $L_{g}h(x)$ is zero due to the vanishing gradient.

Suppose $g_i(x)=c$ is a non-zero constant vector over $\set{X}$. For $n\ge 2$, the null space of $g_i(x)$ is non-empty; there exists $\mu\in\set{R}^n$ such that $\mu^T g_i(x) = \mu^Tc = 0$. The gradient $\nabla h(x)$ along the boundary $\partial \set{C}$ corresponds to the surface normals of the zero-level set. Since $\set{C}$ is compact, $\partial \set{C}$ is compact and the Gauss map $G(x)=\nabla h(x)/||\nabla h(x)||$ is surjective~\cite{le1975complete,alyaseen2025continuity}. There exists at least a point $x\in\partial \set{C}$ such that $\nabla h(x)/||\nabla h(x)||$ is parallel to $\mu$. Since $\mu^Tc =0$, we have $\nabla^T h(x) c/||\nabla h(x)||=0$, and the Lie derivative is $L_{g_i}h(x)=\nabla^T h(x) c=0$. 
\end{proof}

\autoref{insight:cbf_over_compact_sets} generalizes the observations made from~\autoref{ex:standard_cbf}. 
The second point further shows that for cases where $n\ge 2$, if the term $g_i(x)$ is constant, then the $i$-th input will be unconstrained. This case holds for general linear dynamical systems as well as many practical robot systems (e.g., quadrotor systems, wheeled ground vehicles, and manipulators), which is especially problematic in practice as it can lead to unconstrained input right on the safety boundary. 

While related observations were made in~\cite{brunke2024practical,brunke2024preventingunconst, alyaseen2025continuity,black2023adaptation,jankovic2018robust,tan2021high}, we would like to further stress one subtle point that was not formalized in the existing literature: Even if the relative condition is satisfied, the condition in~\eqref{eqn:cbf_qp} can still result in unconstrained scenarios for a particular input channel, as the relative degree condition (\autoref{def:rel-degree}) only requires one of the input dimensions leading to non-zero Lie derivative rather than all the input dimensions. In practice, we do need to ensure that the Lie derivative corresponding to individual input dimensions is active to ensure that all inputs are properly bounded. Thus, we need to ensure is $L_{g_i}h(x) \neq 0$ for all $i\in\set{Z}_{[1,m]}$ rather than just the relative degree being uniformly one over $\set{X}$. \autoref{insight:cbf_over_compact_sets} shows that such points can exist when $g_i(x)$ is a constant vector. We emphasize that similar issues may also arise if $g_i(x)$ is not constant.

Constructing a CBF and subsequently verifying whether its gradient leads to zero Lie derivatives is non-trivial, especially for general nonlinear control-affine systems. In this paper, we propose to rethink the CBF synthesis process by directly designing the gradient field to ensure that the relative degree condition is satisfied by construction. As a result, CBF synthesis can be formulated as boundary value problems with zeroing values along the safety boundaries; the solutions to the boundary value problems can be efficiently obtained via PINNs.

\section{METHODOLOGY}
In this section, we summarize the key results of our proposed PINN-CBF synthesis method. 

\subsection{Sythesizing CBF as Solving Boundary Value Problems}

We first motivate the idea of formulating CBF synthesis as boundary value problems using the simple example we considered in the previous section and then formalize our approach in the subsequent discussion.

\begin{example}[CBF Synthesis as Boundary Value Problems]
Consider again the single integrator example in~\autoref{ex:standard_cbf}. A single continuously differentiable function cannot be used to parameterize the safe set without introducing zero Lie derivatives in the safe set. Suppose we take a gradient-oriented approach and require that $\nabla h_q(x) =\alpha_q g(x) = \alpha_q$ with $|\alpha_q|\neq 0$ for all $x\in \set{X}$, where $q$ is the index of the CBF. Then, by construction, the gradient of the CBF $\nabla h_q(x)$ and hence the Lie derivative $L_gh_q(x)=\nabla h_q(x)$ is non-zero for all $x\in\set{X}$. If we consider two CBFs, each accounting for one extremum of the compact set, then we can formulate two boundary value problems: 
    \begin{align*}
    \nabla h_1 (x) &= \alpha_1 \text{ with }h_1(x_\text{min})=0,\\
    \nabla h_2 (x) &= \alpha_2 \text{ with }h_2(x_\text{max})=0,
\end{align*}
where $\alpha_1 > 0$ and $\alpha_2 < 0$.
For this simple example, the solutions to the boundary value problems can be readily found as $h_1(x) = \alpha_1(x-x_\text{min})$ and $h_2(x) = \alpha_2(x-x_\text{max})$. 
The two CBF candidates result in two input constraints 
    \begin{align*}
    \set{U}_\text{cbf,1}=\{u\in\R^m\:\:|\:\: \alpha_1 u \ge -\gamma_1 \big(\alpha_1(x-x_\text{min})\big)\},\\
    \set{U}_\text{cbf,2}=\{u\in\R^m\:\:|\:\: \alpha_2 u \ge -\gamma_2 \big(\alpha_2(x-x_\text{max})\big)\},
\end{align*} 
where $\gamma_1,\gamma_2\in\K_e$. It is not hard to verify that we can choose $\gamma_1(\xi)=\gamma_2(\xi)=c\:\xi$, a linear function with a positive constant $c$, and show that the constraints imposed by the two CBF candidates result in a non-empty feasible input set for all $x\in\set{X}$. In other words, there exist functions $\gamma_1, \gamma_2\in\K_e$ such that the CBF conditions defined by the two candidates can be satisfied by some input $u$, thereby fulfilling the feasibility condition analogous to the original CBF definition in~\autoref{def:cbf}.  
\end{example}

The above example illustrates how we can instead approach CBF synthesis as a gradient design problem and construct multiple CBFs that jointly cover the desired safe set without encountering issues related to zero Lie derivatives which inactivates the safety filters. This raises several general questions: How should the boundary conditions be defined? How should the gradients be selected, and how can we ensure the feasibility of the resulting safe set? Also, how to solve the boundary value problems for generic control affine systems? We address these questions in the following subsections.

Before introducing our solution, we first present the multi-CBF formulation, which summarizes the key properties that the set of CBFs should satisfy.

\begin{definition}[Multi-CBFs]
\label{def:multi-cbf}
Consider a set of $Q$ continuously differentiable functions $h_q:\set{X}\mapsto\set{R}$. Let $\set{C}_q\subseteq \set{X}$ be the zero-superlevel set of $h_q$ with $\partial\set{C}_q = \{x\in\set{X}\:\:|\:\: h_q(x)=0\}$ and $\nabla h_q(x) \neq 0$ for all $x\in\partial\set{C}_q$. Moreover, $L_{g_i}h_q(x)\neq 0$ for all $i\in\set{Z}_{[1,m]}$, $x\in\set{X}$. The set of functions $(h_1,h_2,...,h_Q)$ are multi-CBFs for \eqref{eqn:nonlinear_affine_control} if there exist $\K_e$-functions $(\gamma_1, \gamma_2,...,\gamma_Q)$ such that
\begin{equation}
\label{eqn:multi_cbf_lie_derivative}
\max_{\u \in \R^m}\min_{q\in\set{Z}_{[1,Q]}}\hspace{0.5em}L_\vec{f} h_q(x) + L_\vec{g} h_q(x) \u + \gamma_q(h_q(x)) \geq 0 
\end{equation}
is satisfied for all $x \in \set{X}$.
\end{definition}

The condition in~\eqref{eqn:multi_cbf_lie_derivative} is analogous to that in~\eqref{eqn:cbf_lie_derivative}, which ensures that there exists at least one control input that satisfies all CBF constraints (i.e., the set of certified inputs is non-empty). 
Given multi-CBFs for system~\eqref{eqn:nonlinear_affine_control}, we can define a set of inputs that satisfy all CBF conditions defined as
\begin{equation}\label{eqn:controller_space_multiple_cbf}
    \set{U}_\text{cbf}(x) = \bigcap_{q=1}^Q \set{U}_{\text{cbf},q}(x),
\end{equation}
where $\set{U}_{\text{cbf},q}(x)=\{u\in\R^m \:\vert \: \dot h_q(x,u) \geq - \gamma_q(h_q(x))\}$ with $\dot h_q(x,u)=L_\vec{f} h_q(x) + L_\vec{g} h_q(x) \u$. We further define 
\begin{equation*}
    \set{C}_Q = \bigcap_{q=1}^Q \set{C}_q.
\end{equation*}

Similar to~\autoref{cor:theorem_2_ames}, if we have a Lipschitz continuous policy $\pi: \set{X} \mapsto \set{R}^m$ such that $\pi(x) \in \set{U}_\text{cbf}(x)$ for all $x \in \set{X}$, where $\set{U}_\text{cbf}$ is defined as in \eqref{eqn:controller_space_multiple_cbf}, then the forward invariance of the sets $\set{C}_q$ and consequently $\set{C}_Q$ can be ensured. In our formulation, given a safe set $\set{C}$, our goal is to construct a set of CBFs that define a set $\set{C}_Q \subseteq \set{C}$. The multi-CBFs can then be used for designing a safety filter to render the safe set safe:
\begin{equation*}
    x(0) \in \set{C}_Q \implies x(t) \in \set{C}, \quad \forall t \in \set{T}_{\x_0}^+.
\end{equation*}

\subsection{Boundary Condition Design}
\label{subsec:boundary_design}
Our goal is to find a set of CBFs to parameterize a set $\set{C}_Q\subseteq \set{C}$ that best preserves the geometry of the safe set $\set{C}$. One of the two important pieces to be designed is the boundary conditions of the CBFs, which determine the boundary of the set to be rendered safe. 

\begin{figure*}
    \centering
    \begin{subfigure}[t]{0.32\textwidth}
        \centering
        \includegraphics[height=0.734375\textwidth]{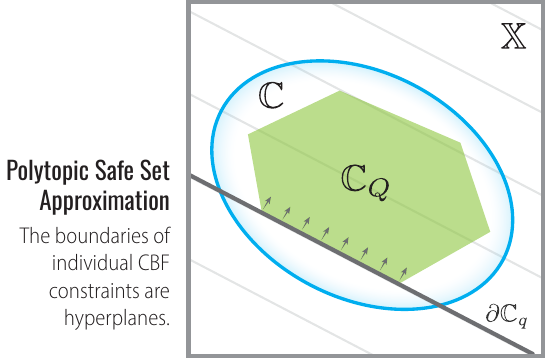}
        \caption{Polytopic safe set approximation~\cite{brunke2024practical}.}
        \label{subfig:polytopics_set}
    \end{subfigure}\hspace{1.5em}
    \begin{subfigure}[t]{0.64\textwidth}
        \centering
        \includegraphics[height=0.3671875\textwidth]{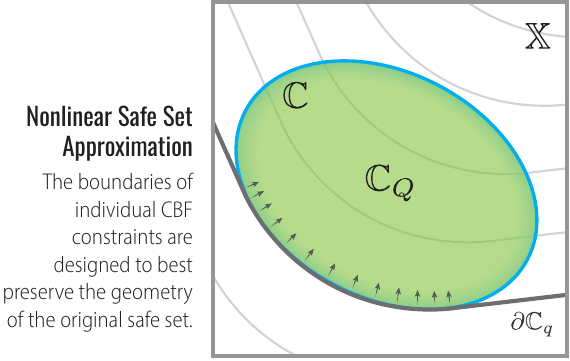}
        \includegraphics[height=0.3671875\textwidth]{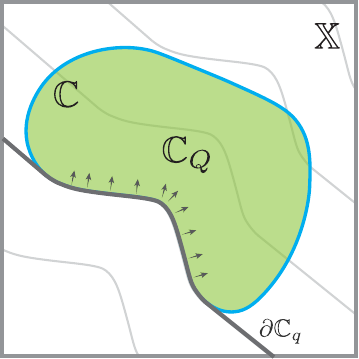}
        \caption{Nonlinear safe set approximation.}
        \label{subfig:nonlinear_set_approximation}
    \end{subfigure}
    \caption{The use of multiple CBFs has been proposed to mitigate issues arising from varying relative degrees. Here, we illustrate the boundaries of approximated safe sets for the approach from~\cite{brunke2024practical} and our proposed method. In these illustrations, the true safe set~$\mathcal{C}$ and the approximated safe set~$\mathcal{C}_Q$ are shown in blue and green, respectively. For each case, the boundary of an individual CBF, $\partial \mathcal{C}_q$, is shown as a dark gray solid line, and the corresponding level sets are illustrated in light gray. The arrows on the boundary $\partial \mathcal{C}_q$ indicate the positive gradient directions. (a)~In~\cite{brunke2024practical}, we proposed using a polytopic set to under-approximate $\set{C}$. In this case, the safety boundaries are hyperplanes. This approach is conservative and is restricted to convex sets. (b)~In this work, we instead use multiple CBFs, where the individual CBFs can have nonlinear boundaries to better preserve the geometry of the original safe set. This approach can be generalized to certain non-convex safe sets where the condition in~\eqref{eqn:surface_normal_condition} can be satisfied for all pairs of points on $\partial \mathcal{C}_q$ for all~$q$.
    }
    \label{fig:set_approximation}
\end{figure*}

In~\cite{brunke2024practical}, we proposed using a polytopic set to under-approximate the safe set in order to mitigate the varying relative degree issue (\autoref{subfig:polytopics_set}). However, such an approximation can be overly conservative and is only applicable to convex sets. To address this, we consider more general half-space-like constraints defined by nonlinear level curves~(\autoref{subfig:nonlinear_set_approximation}). The boundary of the $q$-th CBF candidate is $\partial \set{C}_q=\{x\in\set{X}\:\:|\:\: h_q(x)=0\}$. We require, for any two points $x_i,x_j \in \partial \set{C}_q\cap\partial \set{C}$,
\begin{equation}
\label{eqn:surface_normal_condition}
\nabla^T \hat{h}(x_i)\nabla \hat{h}(x_j)> -1,
\end{equation}
where $\nabla \hat{h}(x)=\nabla^T h(x)/||\nabla h(x)||$ is the normalized gradient vector at a point $x$. This condition allows us to define hypersurfaces with consistent parallel level curves. For convex sets, we can extend the boundary $\partial \set{C}_q\cap\partial \set{C}$ by defining consistent hypersurfaces that satisfy the condition in~\eqref{eqn:surface_normal_condition} for every pair of points in $\partial\set{C}_q$. For non-convex sets, to best preserve the geometry of the safe set, we first compute the convex hull of the safe set, $\text{Conv}(\set{C})$. The set of CBF boundaries is then defined to consist of the disjoint sets in $\partial \set{C} \setminus \left( \partial \text{Conv}(\set{C}) \cap \partial \set{C} \right)$, along with further partitions in $\partial \text{Conv}(\set{C}) \cap \partial \set{C}$. An example is shown in~\autoref{fig:bvp_illustration}.

As discussed in~\autoref{sec:relative_degree}, the varying relative degree issue can arise from two sources: either the gradient of the function $h$ vanishes in the interior of compact sets, or the gradient is orthogonal to $g_i(x)$ at the boundary. The first issue is mitigated by introducing multiple continuously differentiable functions with non-compact zero-superlevel sets. However, the second problem still needs to be addressed. We note that if such a point exists, the original set cannot be rendered safe without encountering the same problem. To address this issue, we consider a local inner approximation of the safe set~(e.g., through the use of single or multiple hyperplanes with constant gradient norms that best match the local geometry while not being orthogonal to $g_i(x)$).

This approach allows us to handle non-convex safe sets as long as we can find a set of partitions where the condition in~\eqref{eqn:surface_normal_condition} can be satisfied for all pairs of points in $\partial \set{C}_q$. We do note that the approach is restricted to connected safe sets, though this is not overly restrictive in practice. If the safe set comprises separate sets, a system initialized within one of the sets will remain within the set. Thus, for a given initial state $x_0\in\set{C}$, at any given time, only the constraints associated with the relevant connected set need to be enforced.

\subsection{Gradient Design}
\label{subsec:gradient_design}
We consider a set of $Q$ CBFs with gradients parameterized in the following form:
\begin{equation}
\label{eqn:gradient_parameterization}
    \nabla h_q(x) =  g(x)\alpha_q(x) +   b(x)\beta_{q}(x),\:\forall q\in\set{Z}_{[1,Q]},
\end{equation}
where $\alpha_q(x) \in \mathbb{R}^{m}$ and $\beta_q(x) \in \mathbb{R}^{n-p(x)}$ are parameters of the target CBF gradient with $p(x)$ denoting the rank of $g(x)$, The matrix $b(x) \in \mathbb{R}^{n \times (n-p(x))}$ has orthonormal columns that span the orthogonal complement of the column space of $g(x)$ such that the columns of $b(x)$ and $g(x)$ together span $\R^n$. 
Note that if $p(x) = n$, we do not require additional vectors to span $\R^{n}$. 
While we keep the formulation general, the rank of $g(x)$ is often constant over $\set{X}$ in practice. 

Each of the CBF candidates defines a nonlinear contour with positive gradient directions pointing toward the interior of the safe set. We can design the gradient of each CBF candidate to ensure that each level set of $h_q(x)$ preserves the shape and orientation of the zero-level set $\partial \set{C}_q$. The union of all the gradients $\nabla h_q(x)$ for all $x \in \partial \set{C}_q$ spans a cone $\set{K}_q$. Let $s_q$ be the vector of length one that corresponds to the central axis of the cone $\set{K}_q$, then we specify the gradient for CBF candidate $h_q$ for all $x \in \set{X}$ as 
\begin{equation*}
    \nabla h_q(x) = \nabla h_q (x_c^*),
\end{equation*}
where $x_c^*$ is the intersection of the boundary $\partial \set{C}_q$ and the set $\set{L} = \{x + \sigma_q s_q\in\set{X}\:\:|\:\: \sigma_q \in \R\}$.

Since the columns of $g(x)$ and $b(x)$ together span $\set{R}^n$, we can parametrize $\nabla h_q(x)$ in the form of~\eqref{eqn:gradient_parameterization}. To ensure that the non-zero Lie derivative condition is satisfied for all $x\in\set{X}$ and the overall input set $\set{U}_\text{cbf}$ is non-empty, we need to introduce additional constraints on the gradient parameters $\alpha_q(x)$ and $\beta_q(x)$. To this end, we define 
\begin{equation}
\label{eqn:gradient_approximation}
    \nabla \tilde{h}_q(x) =  g(x)\tilde{\alpha}_q(x) +   b(x)\tilde{\beta}_{q}(x)
\end{equation}
and formulate an optimization problem to find parameters $\tilde{\alpha}_q(x)$ and $\tilde{\beta}_q(x)$ such that $\nabla \tilde{h}_q(x)$ closely matches the desired target gradient $\nabla h_q(x)$ while ensuring that the additional Lie derivative and feasibility conditions for qualifying $h_q$ as multi-CBFs are satisfied. 

To facilitate the design verification of the feasible input set, we consider a set of parametrized functions $\gamma_q\in\K_e$ with corresponding parameters denoted by $\theta_q$ (e.g., in the linear case $\gamma_q(\xi)=c_q\xi$, the parameter $\theta_q$ is the slope $c_q$). We then formulate the following optimization problem to be solved sequentially for each CBF candidate $q \in \set{Z}_{[1,Q]}$:
\begin{subequations}
\label{eqn:gradient_optimization_full}
\begin{align}
\min_{\tilde{\alpha}_q(x),\tilde{\beta}_q(x),\theta_q} &\:\:||\nabla \tilde{h}_q(x) - \nabla h_q(x)||^2\label{eqn:gradient_optimization1_obj}\\
    \text{s.t.}&\:\: |\tilde{\upsilon}_{q,i}(x)| \ge \epsilon,\:\forall i\in\set{Z}_{[1,m]}\label{eqn:gradient_optimization1_nonnegative}\\
    &\:\: \max_{u\in\set{R}^m}\min_{j\in\set{Z}_{[1,q]}}\: \dot{\tilde{h}}_j(x, u)+\gamma_j (\tilde{h}_j(x))\ge 0,\label{eqn:gradient_optimization1_feasibility}
\end{align}
\end{subequations}
where $\tilde{\upsilon}_{q}(x)=g^T(x)g(x)\:\tilde{\alpha}_{q}(x)$ with $\tilde{\upsilon}_{q,i}(x)$ denoting the $i$-th element of $\tilde{\upsilon}_{q}(x)$, $\epsilon$ is a predefined small positive number, $\nabla \tilde{h}_q(x)$ is parametrized by $(\tilde{\alpha}_q(x), \tilde{\beta}_q(x))$ as defined in \eqref{eqn:gradient_approximation}, and $\dot{\tilde{h}}_j(x, u) =L_f \tilde{h}_j(x) + L_g \tilde{h}_j(x)u$. The objective in~\eqref{eqn:gradient_optimization1_obj} encourages fining parameters $\tilde{\alpha}_q(x)$ and $\tilde{\beta}_q(x)$ that would best match that of the target gradient $\nabla h_q(x)$. The inequality constraints in~\eqref{eqn:gradient_optimization1_nonnegative} ensures that the Lie derivative $L_{g_i} \tilde{h}_q(x)$ is non-zero for all $i\in\set{Z}_{[1,m]}$. 
The constraint in~\eqref{eqn:gradient_optimization1_feasibility} ensures that the constraints constructed up to $q$ will result in a feasible input set (i.e., $\bigcap_{i=1}^q \set{U}_{\text{cbf},i}(x)\neq \emptyset$). The problem in~\eqref{eqn:gradient_optimization_full} is solved sequentially from $q = 1$ to $q = Q$, and the optimal solutions are denoted by $\tilde{\alpha}_q(x)^*$, $\tilde{\beta}_q(x)^*$, and $\theta_q^*$, respectively. After solving the gradient optimization problem for each CBF, we formulate a boundary value problem and solve for $h_q(x)$ following the approach to be discussed in~\autoref{subsec:pinn_solutions}. The solution $h^*_q(x)$ is then used in the gradient optimization of subsequent CBFs.

The constraint in~\eqref{eqn:gradient_optimization1_feasibility} involves a 
min-max problem to ensure that there exists at least one input $u\in\set{R}^m$ that satisfies all CBF constraints up to~$q$. To make the overall optimization problem more tractable, 
we reformulate the gradient optimization problem as
\begin{subequations}
\begin{align}
\min_{\tilde{\alpha}_q(x),\tilde{\beta}_q(x),\atop \theta_q,u} &\:\:||\nabla \tilde{h}_q(x) - \nabla h_q(x)||^2\label{eqn:gradient_optimization2_obj}\\
    \text{s.t.}&\:\: |\tilde{\upsilon}_{q,i}(x)| \ge \epsilon,\:\forall i\in\set{Z}_{[1,m]}\label{eqn:gradient_optimization2_nonnegative}\\
    &\:\:  \dot{\bar{h}}_j(x, u) \ge -\gamma_j (\tilde{h}_j(x)),\:\forall j\in\set{Z}_{[1,q]}\label{eqn:gradient_optimization2_feasibility},
\end{align}
\end{subequations}
where $\dot{\bar{h}}_j(x, u)=L_f\tilde{h}_j(x) + (g(x)\tilde{\alpha}_j(x))^Tg(x)u$ with $\tilde{\alpha}_j(x)=\tilde{\alpha}_j^*(x)$, $\tilde{\beta}_j(x)=\tilde{\beta}_j^*(x)$, $\theta_j=\theta_j^*$, and $\tilde{h}_j(x)=\tilde{h}_j^*(x)$ for all $j\in\{1,2,...,q-1\}$. Note that the constraint in~\eqref{eqn:gradient_optimization2_feasibility} also requires the knowledge of $h_q(x)$. At the boundary $\partial \set{C}_q$, the lower bound in~\eqref{eqn:gradient_optimization2_feasibility} is 0 regardless of the solution $h_q(x)$. For $x\notin \set{C}_q$, we can numerically approximate the value by propagating the gradient from the boundary or by solving the boundary value problem based on the target $\alpha_q(x)$ and $\beta_q(x)$ values.

Suppose $g(x)$ has full column rank, and thus $g^T(x) g(x)$ is invertible. We can simplify the problem by determining the gradient parameters $(\alpha_q(x), \beta_q(x))$ and $\gamma_q$ in two steps. Specifically, we exploit the fact that the column space of $b(x)$ and the column space of $g(x)$ are orthogonal complements and that the columns of $b(x)$ are orthonormal. This leads to the following conditions:
\begin{align*}
        \alpha_q(x) &= \big(g^T(x) g(x) \big)^{-1} g^T(x)\nabla h_q(x),\label{eqn:alpha}\\
    \beta_q(x)&=b^T(x) \nabla h_q(x).
\end{align*}

To ensure that every point satisfies the non-zero Lie derivative condition and that the overall problem leads to a feasible input set, we formulate two optimization problems that are solved in sequence for CBF candidates $q \in\set{Z}_{[1,Q]}$. The first optimization problem finds the best matching $\tilde{\alpha}_q(x)$:
\begin{subequations}
\label{eqn:gradient_optimization_alpha}
\begin{align}
\min_{\tilde{\alpha}_q(x)} &\:\:||\tilde{\alpha}_q(x)- \alpha_q(x)||^2\\
    \text{s.t.}&\:\: |\tilde{\alpha}_{q,i}(x)| \ge \epsilon,\:\forall i\in\set{Z}_{[1,m]},
\end{align}
where $\tilde{\alpha}_{q,i}(x)$ is the $i$-th element of $\tilde{\alpha}_{q}(x)$.
\end{subequations}
The second optimization problem fixes $\tilde{\alpha}_q(x)$ to the optimal solution from~\eqref{eqn:gradient_optimization_alpha} and optimizes $\tilde{\beta}_q(x)$ and $\theta_q$:
\begin{align*}
\min_{\tilde{\beta}_q(x),\theta_q,u} &\:\:||\tilde{\beta}_q(x)- \beta_q(x)||^2\\
     &\:\:  \dot{\bar{h}}_j(x, u) \ge -\gamma_j (\tilde{h}_j(x)),\:\forall j\in\set{Z}_{[1,q]}.
\end{align*}

\subsection{Physics-Informed Neural Solutions}
\label{subsec:pinn_solutions}
Given the boundary conditions and the desired gradients designed in~\autoref{subsec:boundary_design} and \autoref{subsec:gradient_design}, we form the following boundary value problem for the $q$-th CBF:
\begin{subequations}
\label{eqn:boundary_value_problem}
\begin{align}
\nabla \tilde{h}_q(x) &=  g(x)\tilde{\alpha}_q^*(x) +   b(x)\tilde{\beta}_{q}^*(x)\\
\tilde{h}_q(x) &= 0,\:\: \forall x\in\partial\set{C}_q.
\end{align}
\end{subequations}
Solving the problem in~\eqref{eqn:boundary_value_problem} for generic control affine systems and non-affine boundary conditions is not trivial. PINNs are one approach to solving PDEs through a supervised learning approach~\cite{raissi2019physics,baty2024hands}. Our problem in~\eqref{eqn:boundary_value_problem} is a subset of the problems that can be solved through PINNs. A PINN can generally have an architecture that is typical of a neural network. The main difference lies in the training process---a PINN is trained to not only minimize errors in the predicted values but also incorporate terms involving the derivatives of the function approximator to satisfy the underlying physics or differential constraints.

Let $H_q:\set{R}^n\mapsto \set{R}$ denote the function represented by a PINN and $\theta_\text{PINN}$ denote the network parameters. In our case, to solve the boundary value problem in~\eqref{eqn:boundary_value_problem}, we define two loss functions, $E_\text{phy}$ and $E_\text{bc}$, to respectively capture the desired gradient field and the boundary condition:
\begin{align}
\label{eqn:loss_phy}
    E_{\text{phy},q}(\theta_\text{PINN})&=\frac{1}{N_\text{phy}}\sum_{x\in\set{D}_{\text{phy},q}} ||\nabla H_q(x) - \nabla \tilde{h}_q(x)||^2\\
    \label{eqn:loss_bc}
    E_{\text{bc},q}(\theta_\text{PINN})&= \frac{1}{N_\text{bc}}\sum_{x\in\set{D}_{\text{bc},q}} ||H_q(x)- \tilde{h}_q(x)||^2,
\end{align}
where $\set{D}_{\text{phy},q}$ and $\set{D}_{\text{bc},q}$ denote the sets of points used to evaluate the physics and boundary losses, respectively. These can be obtained through sampling from the set~$\set{X}$ and the boundary~$\partial \set{C}_q$. The overall loss function is given by
\begin{align*}
\label{eqn:loss_function}
    E_q(\theta_\text{PINN}) 
    = \lambda E_{\text{bc},q}(\theta_\text{PINN}) + E_{\text{phy},q}(\theta_\text{PINN}),
\end{align*}
where $\lambda\in\set{R}_{>0}$ is a hyperparameter that weights the relative importance of the two loss terms and can be either predefined or automatically tuned~\cite{baty2024hands}.

In our work, we use a fully connected feedforward network architecture with continuously differentiable activation functions (e.g., hyperbolic tangent). For a network with a finite depth of $D$, the function $H_q$ can be expressed as
\begin{equation}
\label{eqn:net_architecture}
    H_q(x) = l_{D-1} \circ n_{D-1}  \circ \cdots \circ n_2 \circ l_1 \circ n_1\circ l_0 (x),
\end{equation}
where $n_i:\set{R}^{N_i}\mapsto\set{R}^{N_i}$ and $l_i:\set{R}^{N_{i-1}}\mapsto\set{R}^{N_i}$ represent the nonlinear activation layers and linear transformation layers, respectively, with $N_i$ being the width of layer~$i$.

Note that we accounted for the boundary conditions by introducing the corresponding loss function~\eqref{eqn:loss_bc} in the training process. We can alternatively enforce the boundary conditions exactly through the architecture of the PINN if the boundary condition can be expressed in terms of a well-behaved trial function $B$ that satisfies the boundary condition by construction~\cite{lagaris1998artificial}. In this case, we define the PINN as
\begin{equation*}
   H_q(x) =  B \circ \phi(x),
\end{equation*}
where $\phi(x) = l_{D-1} \circ n_{D-1}  \circ \cdots \circ n_2 \circ l_1 \circ n_1\circ l_0 (x)$ as in~\eqref{eqn:net_architecture}, and $B(\xi)= F(x)\xi$ with $F(x)=0$ for $x\in \partial \set{C}_q$. Thus, through this approach, by construction, we can enforce PINN to have zero output for $x\in \partial \set{C}_q$ as required by our CBF formulation. The remaining process then follows the standard PINN training procedure.

\section{SIMULATION AND EXPERIMENTAL RESULTS}

We evaluate our PINN-based CBF synthesis using a quadrotor in simulation and the real world. 
A video of the experiments can be seen at this link: \href{http://tiny.cc/relative-degree-pinn-cbf}{http://tiny.cc/relative-degree-pinn-cbf}. 

In both examples, we use a non-linear feedback controller~\cite{mellinger2011minimum} to stabilize the quadrotor's attitude and constrain its motion to the $z$-axis. The system dynamics along the $z$-axis are given by
\begin{equation*}
    \dot{x} = \begin{bmatrix}
        \dot{z} &
        k_1 - \text{g} + k_2 u
    \end{bmatrix}^T,
\end{equation*}
where $x = [z, \dot{z}]^T \in \R^2$ is the state, $u \in \R$ is the mass-normalized acceleration input, $\text{g} = 9.81$ m/s$^2$ is the gravitational constant, and $k_1, k_2 \in \R$ are the system parameters. In the simulation, $k_1 = 20.91$ and $k_2 = 3.65$ and in the real world $k_1 = 20.91$ and $k_2 = 2.19$. In this example, $g(x)$ is constant over the state space and yields $L_g h(x) = 0$ for two states on the boundary of the safe set $\set{C}$. 
For both settings, each PINN has three hidden layers with 100 neurons and uses the $\tanh$ activation function. 
We run the safety filter at \SI{100}{\hertz}~in the simulation and \SI{70}{\hertz}~in the real world. 

\begin{figure}
    \centering
    \input{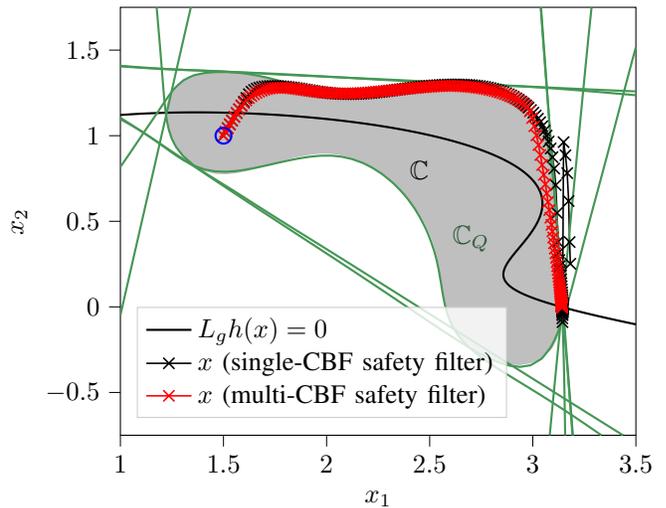}
    \caption{Simulation results of our proposed PINN-based CBF synthesis approach for rendering a non-convex set safe. We can accurately approximate the desired safe set~(gray shaded area) using ten PINNs~(zero-level sets in green). We compare the single-CBF and our proposed multi-CBF approach starting from the same initial condition~(blue circle). The system's state in closed-loop with the single-CBF safety filter~(black crosses) violates the desired safety constraint. In contrast, the state of the system in closed-loop with the CBF safety filter using our PINN-based CBFs~(red crosses) stays inside the safe set for all future time, although the system converges to a state where $L_g h(x) = 0$ for the single CBF (black solid line).}
    \label{fig:non-convex-example}
    \vspace{-1em}
\end{figure}

We consider different safe sets in simulation and in real-world experiments, respectively, illustrating the applicability of our approach to a non-convex and a convex set. In the simulation, the desired safe set is derived from a continuously differentiable Himmelblau function~\cite{himmelblau1972applied}. We approximate the non-convex safe set using ten PINNs.  
We compare the standard single-CBF safety filter approach directly using the Himmelblau function as the CBF with our proposed multi-CBF safety filter that synthesizes multiple CBFs based on the Himmelblau function using PINNs while also accounting for the varying relative degree.   
The closed-loop trajectories of both safety filters are initialized at the same state~(blue circle), and the same unsafe control input trajectory $\pi(x) = 0.4$~m/s$^2$ is applied to both systems for the duration of the simulation. As shown in~\autoref{fig:non-convex-example}, the single-CBF safety filter leads to constraint violations caused by reaching the set of states where $L_g h(x) = 0$ intersects the boundary of $\set{C}$. Our proposed multi-CBF safety filter also converges to the same states, but by design $L_g h_q(x) \neq 0$ for all PINN-based CBFs, and consequently, the safety filter renders~$\set{C}_Q\subseteq \set{C}$ safe without chattering issues or constraint violations. 

\begin{figure*}
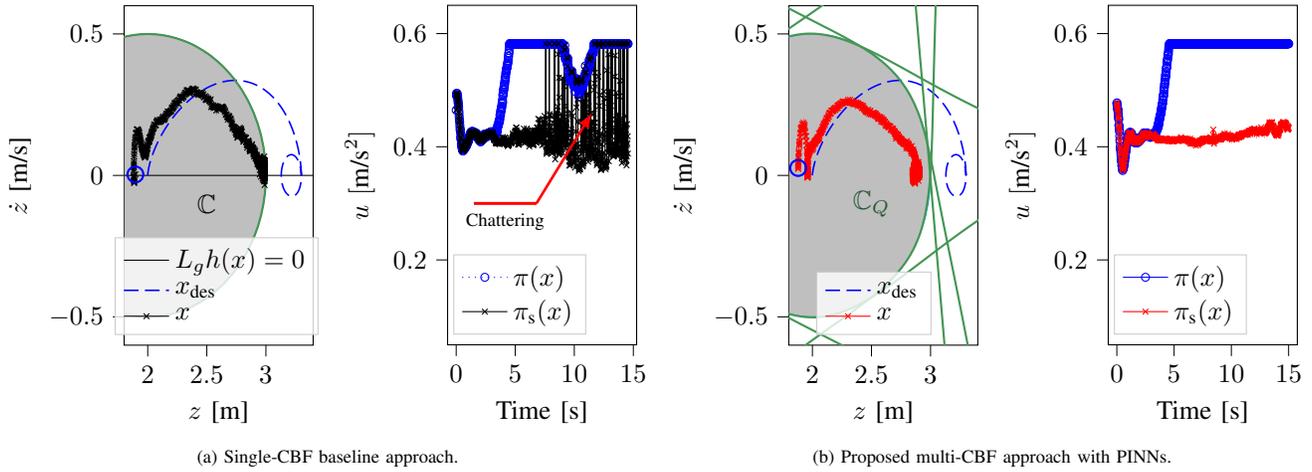

    \centering
        \centering
    \begin{subfigure}[t]{0.49\textwidth}
        \centering
    \input{figures/drone_no_pinn_state}
    \input{figures/drone_no_pinn_thrust}
    \vspace{-0.42cm}
    \caption{Single-CBF baseline approach.}
    \end{subfigure}
    \begin{subfigure}[t]{0.49\textwidth}
        \centering
        \input{figures/drone_pinn_state}
        \input{figures/drone_pinn_thrust}
        \caption{Proposed multi-CBF approach with PINNs.}
    \end{subfigure}
    \caption{Experimental results of our proposed PINN-based CBF synthesis approach for rendering a convex set safe. Similar to the simulation example, we approximate the desired safe set (gray shaded area) using eight PINNs (zero-level sets in green). The single-CBF baseline uses a quadratic CBF, which results in a set where $L_g h(x) = 0$ inside of $\set{C}$ (black solid line). The quadrotor is initialized at similar states (blue circle), and an uncertified policy~(blue dashed line) is used to drive the quadrotor from the interior of the safe set to the unsafe region. With the standard single-CBF safety filter, as the quadrotor approaches the safety boundary~(black crosses in the first panel), we observe large input oscillations near the boundary of the safe set due to filter inactivity~(second panel). In contrast, with our proposed multi-CBF method, as the quadrotor approaches the safety boundary~(red crosses in the third panel), the input oscillations are mitigated through proper CBF gradient design~(fourth panel).
    }
    \label{fig:drone_experiment}
    \vspace{-1.5em}
\end{figure*}

In the real-world example, the safe set is represented as an ellipsoid defined by a quadratic function. The uncertified controller $\pi(x)$ is tasked to drive the quadrotor from a point inside the safe set to a point outside of it. As shown in \autoref{fig:drone_experiment}, with the single-CBF approach where the quadratic function is directly used as the CBF, the quadrotor experiences significant input chattering as it approaches the boundary of the safe set, especially in regions close to the set where the CBF becomes inactive (i.e., where $L_g h = 0$). In contrast, with our proposed multi-CBF approach, with properly designed CBF gradients, the quadrotor remains within the safe set without exhibiting any chattering behaviour in the filtered input command. This set of experiments further illustrates the efficacy of our approach in mitigating the adverse effects caused by varying relative degrees in practical systems.

\section{CONCLUSIONS}
In this work, we proposed an alternative view on control barrier function (CBF) synthesis to address the issue of varying relative degrees. In particular, we design CBFs by first specifying their gradient fields and formulating CBF synthesis as boundary value problems, which are solved using physics-informed neural networks (PINNs). This approach allows us to mitigate varying relative degrees and the resulting CBF inactivity issues without relying on conservative safe set approximations or retrospective verification and modification. Through quadrotor simulations and real-world experiments, we demonstrate that our approach successfully mitigates the chattering behaviour caused by inactive safety filters and thereby enables the certification of systems with the desired safety guarantees.

\addtolength{\textheight}{-3cm}   


\balance

\bibliographystyle{IEEEtran}
\bibliography{IEEEabrv, references}

\balance
\end{document}